\documentclass[reprint,nofootinbib,amsmath,amssymb,aps]{revtex4-2}

\pdfoutput=1
\usepackage[pdftex]{graphicx}
\usepackage[pdftex]{xcolor}
\usepackage[colorlinks,bookmarks]{hyperref}
\definecolor{linkblue}{rgb}{0,0,0.8}
\definecolor{linkgreen}{rgb}{0,0.5,0}
\hypersetup{linkcolor=linkblue, citecolor=linkgreen, urlcolor=linkblue}

\usepackage[T1]{fontenc}
\usepackage[utf8]{inputenc}
\usepackage{lmodern}
\usepackage[english]{babel}

\usepackage{dcolumn}
\usepackage{bm}
\usepackage{xcolor} 
\usepackage{booktabs} 
\usepackage[normalem]{ulem}
\usepackage{multirow}
\usepackage{makecell}

\definecolor{valecol}{rgb}{0,0.5, 1.}

\graphicspath{{./figs/}}

\begin{document}

\preprint{...}

\title{Dark Degeneracy in DESI DR2: Interacting or Evolving Dark Energy?}

\author{Vitor Petri$^{1}$}\email{vitorpetrisilva@gmail.com}
\author{Valerio Marra$^{2,3,4}$}
\author{Rodrigo von Marttens$^{5}$}
\affiliation{$^{1}$PPGFis, Universidade Federal do Espírito Santo, 29075-910, Vitória, ES, Brazil\\
$^{2}$Departamento de Física, Universidade Federal do Espírito Santo, 29075-910, Vitória, ES, Brazil\\
$^{3}$INAF -- Osservatorio Astronomico di Trieste, via Tiepolo 11, 34131, Trieste, Italy\\
$^{4}$IFPU -- Institute for Fundamental Physics of the Universe, via Beirut 2, 34151, Trieste, Italy\\
$^{5}$Instituto de Física, Universidade Federal da Bahia, 40210-340, Salvador, BA, Brazil}

\date{\today}

\begin{abstract}
The standard $\Lambda$CDM model, despite its success, is challenged by persistent observational tensions in the Hubble constant ($H_0$) and the matter clustering amplitude ($S_8$), motivating the exploration of alternative cosmological scenarios.
We investigate a dark energy model with a phenomenological interaction in the dark sector, constructed to be exactly degenerate at the background level with the Chevallier-Polarski-Linder (CPL) parameterization. This setup allows us to test whether models with identical expansion histories but distinct physical mechanisms can be distinguished by cosmological data. We perform a Bayesian analysis using a combination of recent datasets: DESI DR2 BAO measurements, DESY5 supernovae, and CMB data from \textit{Planck} and ACT. We find that both the interacting model and the CPL model provide significantly better fits to the data than $\Lambda$CDM.
Although indistinguishable in background observables, the interacting model predicts a distinct matter-sector evolution driven by a late-time sign change in the dark sector interaction at $z \approx 0.8$, corresponding to the $w=-1$ crossing in the CPL description. In this sense, the interacting picture may be considered more physical, since it avoids the problematic crossing by construction.
The resulting decay of dark energy into dark matter lowers $S_8$, potentially alleviating the weak-lensing $S_8$ tension. At the same time, it predicts a sharp suppression of the growth rate $f\sigma_8(z)$ at $z \lesssim 0.8$, which is in tension with current measurements of structure formation. This indicates that the model may not simultaneously reconcile the expansion history and the observed growth of cosmic structure, highlighting the need for a more comprehensive analysis to fully assess its viability.
\end{abstract}

\maketitle

\section{Introduction}

The standard model of cosmology, the flat $\Lambda$ Cold Dark Matter ($\Lambda$CDM) model, has achieved remarkable success over the past decades. It provides a robust description of a wide range of cosmological observations, from the temperature and polarization anisotropies in the Cosmic Microwave Background (CMB) to the large-scale distribution of galaxies \citep{Planck:2018vyg}. Despite its successes, the $\Lambda$CDM paradigm rests on two pillars of unknown physical nature: Cold Dark Matter (CDM) and, most enigmatically, Dark Energy (DE), typically modeled as a cosmological constant, $\Lambda$.

The cosmological constant interpretation of dark energy, while consistent with most data, is plagued by profound theoretical challenges, such as the fine-tuning and cosmic coincidence problems \citep{Weinberg:1988cp,Velten:2014nra}. This has motivated a vast theoretical effort to explore alternative explanations, proposing that dark energy is a dynamical field or a manifestation of modified gravity. Furthermore, in recent years, the success of $\Lambda$CDM has been challenged by a series of persistent tensions between different cosmological probes \citep{CosmoVerseNetwork:2025alb}. The most significant of these are the Hubble tension, a $\sim 5\sigma$ discrepancy between the value of the Hubble constant $H_0$ measured locally from Cepheids and Type Ia Supernovae (SNe Ia) and that inferred from CMB data \citep{Riess:2021jrx, Planck:2018vyg}, and the $S_8$ tension, where the amplitude of matter clustering $S_8 = \sigma_8\sqrt{\Omega_m/0.3}$ measured by weak lensing surveys is consistently lower than the CMB-inferred value \citep{DES:2021bvc, Heymans:2020gsg}.

These tensions have intensified the search for new physics beyond the standard model. In this context, we have entered a new era of precision cosmology, driven by next-generation galaxy surveys. The recent Data Release 2 (DR2) from the Dark Energy Spectroscopic Instrument (DESI) has provided the most precise measurements of Baryon Acoustic Oscillations (BAO) to date, mapping the expansion history of the Universe with unprecedented accuracy across a vast cosmic volume \citep{DESI:2025zgx}. This wealth of high-quality data provides a powerful new arena in which to test alternative cosmological models.

In this paper, we explore a potential solution to the dark energy puzzle by investigating a model with a phenomenological interaction in the dark sector. Specifically, we study a model where the interaction is constructed to be degenerate at the background level with the widely-used Chevallier-Polarski-Linder (CPL) parameterization \citep{Chevallier:2000qy,Linder:2002et} for a dynamical dark energy fluid. This model choice is motivated by recent results that indicate a preference for evolving dark energy within the CPL framework \citep{DESI:2025zgx}. This setup provides a unique opportunity to test whether two physically distinct mechanisms, which produce the exact same cosmic expansion history, can be distinguished by the latest cosmological data. We test this model against a comprehensive dataset including the latest results from DESI, \textit{Planck}, ACT, and the DESY5 supernova sample.

This paper is organized as follows. In Section~\ref{sec:model}, we present the theoretical framework for both the interacting model \citep{amendola2010dark} and its degenerate CPL counterpart. In Section~\ref{sec:data}, we describe the datasets and likelihoods used in our analysis. We present our main findings in Section~\ref{sec:results}, detailing the parameter constraints, the physical interpretation of the model's behavior, the statistical model comparison, and the impact on cosmological tensions. Finally, we summarize our findings and discuss their implications in Section~\ref{sec:conclusion}.

\section{Theoretical Framework}\label{sec:model}

The theoretical framework of this work is based on a model that explores a degeneracy within the dark sector, allowing a dynamical dark energy model to be mapped into an interacting dark sector model. We begin by outlining the general background cosmology.

In a homogeneous and isotropic Universe, described by the Friedmann-Lemaître-Robertson-Walker (FLRW) metric, the evolution of non-interacting radiation (r) and baryons (b) is governed by their respective conservation equations:
\begin{align}
\dot{\rho}_r+4H\rho_r=0 \quad \Rightarrow \quad \rho_r = \rho_{r0}\,a^{-4} \,, \\
\dot{\rho}_b+3H\rho_b=0 \quad \Rightarrow \quad \rho_b = \rho_{b0}\,a^{-3} \,, \label{eq:rhob} 
\end{align}
where an overdot denotes differentiation with respect to cosmic time $t$, and $H$ is the Hubble rate satisfying $3H^{2} = 8\pi G\rho$, with $\rho$ the total energy density.
The present-day density parameter for the $i$-th component is $\Omega_{i0} = 8\pi G\rho_{i0}/3H_0^2$.

Following \citet{vonMarttens:2019ixw}, the dark sector, comprising cold dark matter (subscript $c$) and dark energy (subscript $x$), can be treated as a single, unified dark fluid (subscript $d$). Its energy density $\rho_d$ and pressure $p_d$ are given by:
\begin{align}
    \rho_d &= \rho_c+\rho_x, \label{rho_d} \\
    p_d &= p_x = w_x(a)\rho_x, \label{p_d}
\end{align}
where we allow the DE equation of state (EoS) parameter, $w_x$, to be a function of the scale factor $a$.
It is convenient to define the ratio of the dark sector energy densities, $r(a) \equiv \rho_c/\rho_x$. Using this ratio, we can define an effective EoS for the unified dark fluid, $p_d = w_d(a)\rho_d$, where:
\begin{equation}\label{w_d(a)}
    w_d(a) = \frac{w_x(a)}{1+r(a)}.
\end{equation}
The unified dark fluid must obey the standard conservation equation:
\begin{equation}
    \dot{\rho}_d+3H[1+w_d(a)]\rho_d = 0,
\end{equation}
which has the formal solution:
\begin{equation}\label{rho_d(a)}
    \rho_d(a) = \frac{3H^2_0}{8\pi G}\Omega_{d0}\exp\left[-3\int_1^a \frac{1+w_d(a')}{a'}da'\right].
\end{equation}

The cornerstone of this model is the degeneracy apparent in Eq.~\eqref{w_d(a)}: different combinations of $w_x(a)$ and $r(a)$ can produce the exact same effective EoS $w_d(a)$, leading to an identical Hubble expansion history $H(z)$. We explore this degeneracy by considering two physically distinct scenarios that yield the same background cosmology. We denote quantities related to the dynamical DE approach with a bar (e.g., $\bar{\rho}_c$) and those related to the interacting approach with a tilde (e.g., $\tilde{\rho}_c$).

\subsection{Dynamical DE Model}
In this scenario, CDM and DE are separately conserved, but DE has a time-varying EoS, $\bar{w}_x(a)$. The conservation equations are:
\begin{align}
    &\dot{\bar{\rho}}_c + 3H\bar{\rho}_c = 0 \quad \Rightarrow \quad \bar{\rho}_c = \bar{\rho}_{c0}a^{-3}, \label{bar rho c} \\
    &\dot{\bar{\rho}}_x + 3H\bar{\rho}_x[1+\bar{w}_x(a)] = 0 \nonumber\\
    &\quad \Rightarrow \quad \bar{\rho}_x = \bar{\rho}_{x0}\exp\left[-3\int_1^a\frac{1+\bar{w}_x(a')}{a'}da'\right]. \label{bar rho x}
\end{align}
The ratio $\bar{r}(a) = \bar{\rho}_c / \bar{\rho}_x$ is then given by:
\begin{equation}\label{bar r}
    \bar{r}(a) = \frac{\bar{\Omega}_{c0}}{\bar{\Omega}_{x0}} a^{-3} \exp\left[+3\int_1^a \frac{1+\bar{w}_x(a')}{a'}da'\right], 
\end{equation}
and the effective EoS of the dark sector is $w_d(a) = \bar{w}_x(a) / [1+\bar{r}(a)]$.

\subsection{Interacting DE Model}
In this second scenario, the DE EoS is fixed to that of a cosmological constant, $\tilde{w}_x = -1$, but CDM and DE exchange energy. The conservation equations become:
\begin{align}
    \dot{\tilde{\rho}}_{c}+3H\tilde{\rho}_{c} &= \tilde{Q}, \label{cdmenergyint} \\
    \dot{\tilde{\rho}}_{x} &= -\tilde{Q}, \label{deenergyint}
\end{align}
where $\tilde{Q}$ is the interaction term. Following \citet{vonMarttens:2019ixw}, the interaction can be characterized by the function $\tilde{f}(\tilde{r})$:
\begin{equation}\label{EDO tilde r}
    \dot{\tilde{r}}+3H\tilde{r}\big[\tilde{f}\left(\tilde{r}\right)+1\big]=0,
\end{equation}
where $\tilde{f}(\tilde{r})$ is related to the interaction term $\tilde{Q}$ via the following definition,
\begin{equation}\label{f tilde def}
    \tilde{f}\left(\tilde{r}\right) \equiv \frac{\tilde{Q}}{3H}\left(\frac{\tilde \rho_c+ \tilde \rho_x}{\tilde \rho_c\, \tilde \rho_x}\right),
\end{equation}
or, equivalently, can be determined from the evolution of $\tilde{r}(a)$ itself:
\begin{equation}\label{f tilde}
    \tilde{f}\left(\tilde{r}\right) = -\frac{a\tilde{r}'}{3\tilde{r}}-1,
\end{equation}
with the prime denoting a derivative with respect to the scale factor $a$.

\subsection{Equivalence Description}
To establish the degeneracy, we enforce that the effective dark EoS, $w_d(a)$, is identical in both models. By equating $w_d$ from the dynamical model with that from the interacting model (where $w_d = -1/[1+\tilde{r}(a)]$), we find the mapping:
\begin{equation} \label{rtilde_map}
    \tilde{r}\left(a\right) = -\frac{1+\bar{r}\left(a\right)}{\bar{w}_{x}\left(a\right)} - 1.
\end{equation}
This equation allows any dynamical model specified by $\bar{w}_x(a)$ to be mapped to an equivalent interacting model described by $\tilde{r}(a)$, which produces the same cosmic expansion. Furthermore, the individual dark sector densities can also be mapped:
\begin{align}
    \tilde{\rho}_{c}(a) &= \bar{\rho}_{c}(a) + \bar{\rho}_{x}(a)\left[1+\bar{w}_{x}(a)\right], \label{rhoc2} \\
    \tilde{\rho}_{x}(a) &= -\bar{w}_{x}(a)\bar{\rho}_{x}(a). \label{rhox2}
\end{align}

These relations show that while the energy density of the unified dark fluid $\rho_d$ is the same in both pictures, the constituent densities $(\tilde{\rho}_c, \tilde{\rho}_x)$ and $(\bar{\rho}_c, \bar{\rho}_x)$ evolve differently. Consequently, their present-day density parameters will differ ($\bar{\Omega}_{c0} \neq \tilde{\Omega}_{c0}$, $\bar{\Omega}_{x0} \neq \tilde{\Omega}_{x0}$).

From a theoretical perspective, this arises because Einstein’s equations constrain only the total energy–momentum tensor, without fixing the contributions of its individual components. From an observational perspective, although the Hubble rate is identical in the two scenarios, the corresponding values of the cosmological parameters differ, and this distinction plays a central role in the statistical analysis presented in this paper.

\subsection{Linear Perturbation Framework}

With the background-level degeneracy established, we now define the perturbative framework for our analysis. While the formalism developed in \citet{vonMarttens:2019ixw} allows for a complete degeneracy at the linear level under specific conditions regarding sound speed and anisotropic shear, our goal is to explore how models that are identical in their expansion history can be distinguished by their impact on structure formation.

To do this, we deliberately break the full degeneracy by making a common, physically-motivated assumption for both the dynamical and interacting models: we set the comoving sound speed of dark energy to unity ($\bar{c}_{s}^{2} = \tilde{c}_{s}^{2} = 1$). This choice is characteristic of standard scalar field models (quintessence) \citep{amendola2010dark}.

For the interacting model, we further specify the nature of the perturbative interaction. We assume no perturbation in the energy transfer rate ($\delta\tilde{Q} = 0$) and that there is no momentum transfer in the CDM rest frame. This latter condition provides a simple relation for the momentum transfer potential $\mathcal{\tilde{F}}$ in terms of the velocity divergences of the total fluid ($\theta$) and the CDM ($\theta_c$), such that $k^{2}\mathcal{\tilde{F}} = \tilde{\rho}_c(\theta - \theta_c)$.

The evolution of linear perturbations is governed by the conservation of energy and momentum. In the Newtonian gauge, it is convenient to write these equations using the density contrast, $\delta_{i} \equiv \delta\rho_{i}/\rho_{i}$, and the comoving sound speed, $c_{s(i)}^{2} \equiv \delta p_{i}^{(c)}/\delta\rho_{i}^{(c)}$. The general gauge-invariant equations for the evolution of the density contrast ($\delta_i'$) and the velocity divergence ($\theta_i'$) for each component $i$ are given by \citep{Ma:1995ey,Valiviita:2008iv}:

\begin{align}
&\delta_i' + 3 \mathcal{H} ( c_{s(i)}^2 - w_i ) \delta_i + (1 + w_i)( \theta_i + 3\Phi' ) \nonumber\\
&+ 9\mathcal{H}^2 (1 + w_i) ( c_{s(i)}^2 - w_i ) \frac{\theta_i}{k^2}= \nonumber \\
&\frac{a Q_i}{\rho_i} \left[ \Psi - \delta_i + 3 \mathcal{H} ( c_{s(i)}^2 - w_i ) \frac{\theta_i}{k^2} \right] + \frac{a \delta Q_i}{\rho_i}, \label{eq:delta_prime}
\end{align}
and for a fluid with $w_i \neq -1$:
\begin{align}
&\theta_i' + \mathcal{H} ( 1 - 3 c_{s(i)}^2 ) \theta_i - \frac{c_{s(i)}^2}{1 + w_i} k^2 \delta_i - k^2 \Psi = \nonumber\\
&\frac{a Q_i}{\rho_i (1 + w_i)} \left[ \theta - ( 1 + c_{s(i)}^2 ) \theta_i \right] - \frac{a k^2 \mathcal{F}_i}{\rho_i(1+w_i)}. \label{eq:theta_prime}
\end{align}

In these equations, a prime denotes a derivative with respect to conformal time, $\mathcal{H}$ is the conformal Hubble rate, and $\Psi$ and $\Phi$ are the metric potentials in Newtonian gauge. For the non-interacting dynamical DE model, all interaction terms on the right-hand side ($Q_i, \delta Q_i, \mathcal{F}_i$) are set to zero. For the interacting model, since $\tilde{w}_x = -1$, Eq. \eqref{eq:theta_prime} is not used for the DE component; its velocity divergence $\tilde{\theta}_x$ is considered non-dynamical and can be set to zero.

The conservation equations~\eqref{eq:delta_prime} and \eqref{eq:theta_prime} govern the evolution of the fluid perturbations. They form a coupled system with the perturbed Einstein equations, which determine the evolution of the metric potentials $\Phi$ and $\Psi$. The total perturbed energy-momentum tensor, which sources the Einstein equations, includes the contributions from all fluids and is modified by the dark sector interaction terms ($Q_i, \delta Q_i, \mathcal{F}_i$) in the interacting scenario.
This modified source term leads to a different evolution for $\Phi$ and $\Psi$ compared to the non-interacting CPL model, even when the background expansion is identical. These metric potentials are, in turn, the fundamental quantities used to compute all cosmological observables related to perturbations.
Specifically, the CMB power spectra are calculated by evolving the full photon-baryon Boltzmann hierarchy, which is directly coupled to $\Phi$ and $\Psi$ (e.g., via the Sachs-Wolfe and integrated Sachs-Wolfe effects). Similarly, the lensing potential is defined by the line-of-sight integral of the sum of the metric potentials, $\Phi+\Psi$. In this way, the interacting model generates unique predictions for observables sensitive to the growth of structure, such as the CMB lensing spectrum and the late-time ISW effect, allowing it to be distinguished from its degenerate CPL counterpart.

\subsection{The CPL Parameterization}

The specific dynamical DE model adopted in this work is the CPL parameterization. It is characterized by a time-varying EoS of the form:
\begin{equation}
    \bar{w}_{x}(a) = w_{0} + w_{a}(1-a).
\end{equation}
This model provides a simple, two-parameter description of dynamical DE and reduces to the standard $\Lambda$CDM model when $w_0 = -1$ and $w_a = 0$.

For this parameterization, the energy densities in the dynamical (non-interacting) description are given by:
\begin{align}
    \bar{\rho}_{c}(a) &= \bar{\rho}_{c0}a^{-3}, \\
    \bar{\rho}_{x}(a) &= \bar{\rho}_{x0}a^{-3(1+w_{0}+w_{a})}\exp\left[3w_{a}(a-1)\right].
\end{align}
Using the mapping relations from Eqs.~\eqref{rhoc2} and \eqref{rhox2}, the corresponding energy densities in the equivalent interacting model are:
\begin{align}
    \tilde{\rho}_{c}(a) = \dfrac{3H_{0}^{2}}{8\pi G}a^{-3} \lbrace &\bar{\Omega}_{c0} + \bar{\Omega}_{x0}\left[1+w_{0}+w_{a}(1-a)\right]\nonumber \\
    &a^{-3(w_{0}+w_{a})}\exp\left[-3w_{a}(1-a)\right] \rbrace, \label{rhoc2cpl}
\end{align}
\begin{align}
    \tilde{\rho}_{x}(a) = -\dfrac{3H_{0}^{2}}{8\pi G}\bar{\Omega}_{x0}[&w_{0}+w_{a}(1-a)] a^{-3(1+w_{0}+w_{a})}\nonumber\\
    &\exp\left[-3w_{a}(1-a)\right]. \label{rhox2cpl}
\end{align}
From these densities, the ratio $r \equiv \rho_c/\rho_x$ in each framework can be found. For the interacting model, $\tilde{r}(a)$ is explicitly given by the mapping:
\begin{equation}\label{rcpl_tilde}
    \tilde{r}(a) = -\dfrac{1+\bar{r}(a)}{w_{0}+w_{a}(1-a)} - 1,
\end{equation}
where $\bar{r}(a) = \bar{r}_{0}\exp\left[3w_{a}(1-a)\right]a^{3(w_{0}+w_{a})}$. The interaction function $\tilde{f}(\tilde{r})$ that generates this evolution is derived from Eq.~\eqref{f tilde} and has the following complex form:
\begin{align}
\tilde{f}(\tilde{r}) &= \{ 3w_{0}(1+w_{0}) + w_{a}[3+6w_{0}-2a(1+3w_{0})] \nonumber\\
&+ 3w_{a}^{2}(1-a)^{2}\} \left\{ \bar{r}_{0}a^{3(w_{0}+w_{a})} + \exp\left[-3w_{a}(1-a)\right] \right\}\nonumber\\
&\lbrace3\left[w_{0}+w_{a}(1-a)\right]\{ \bar{r}_{0}a^{3(w_{0}+w_{a})} \nonumber\\
&+\! \left[1+w_{0}+w_{a}(1\!-\!a)\right]\exp\left[-3w_{a}(1-a)\right] \}\rbrace^{-1}\label{frcpl} \!.
\end{align}

A key feature of this interaction is its complexity, arising from the two free parameters $w_0$ and $w_a$. Depending on their values, the sign of $\tilde{f}(\tilde{r})$ can change over cosmic time, implying a change in the direction of energy transfer between dark matter and dark energy.

An essential requirement of this framework is that the interaction must become negligible in the early Universe, ensuring that the standard $\Lambda$CDM evolution is recovered at high redshifts.  
By construction, our IDE model is exactly background-degenerate with a CPL dark energy model, implying that both share the same expansion history for any scale factor.  
In practice, the CPL parameters favored by current data produce late-time deviations from $\Lambda$CDM, while the early Universe remains effectively indistinguishable from it, see Figure~\ref{fig:omegas}.  
In other words, the interaction term is dynamically irrelevant at early times and only becomes significant in the recent Universe, driving the departures from $\Lambda$CDM observed at low redshift.

\section{Datasets and Likelihoods}\label{sec:data}

To constrain the theoretical models, we perform a Bayesian statistical analysis using the cosmological inference code \texttt{Cobaya} \citep{Torrado:2020dgo}. The theoretical predictions, including the evolution of background and perturbative quantities, are computed with a modified version of the Einstein-Boltzmann solver \texttt{CLASS} \citep{Blas:2011rf}. We sample the posterior distributions of the parameters using the Monte Carlo Markov Chain (MCMC) method implemented in \texttt{Cobaya}. The datasets used in our analysis are described below. Codes and related files are available at \citep{vitor_petri_2025_17977651}.

\subsection{Cosmic Microwave Background}

The power spectra of temperature and polarization anisotropies in the CMB provide a cornerstone for modern cosmological parameter inference. Our analysis uses the most recent data from the \textit{Planck} satellite, complemented by other experiments.

For the primary anisotropies, our baseline analysis is built upon the \textit{Planck} Legacy 2018 data, specifically using the latest PR4 data processing release (\texttt{NPIPE}) \citep{Rosenberg:2022sdy}. We use a combination of likelihoods: for angular scales at low multipoles ($\ell < 30$), we employ the \texttt{simall} and \texttt{Commander} likelihoods for temperature and polarization; for high multipoles ($\ell \geq 30$), we use the \texttt{CamSpec} likelihood for the temperature (TT), polarization (EE), and cross-correlation (TE) power spectra \citep{Wilding:2020oza, Rosenberg:2022sdy}. This approach replaces the older \texttt{Plik} likelihood from the PR3 release and is known to mitigate the so-called $A_L$ lensing anomaly present in previous data releases.

In addition to the primary anisotropy spectra, we incorporate the power spectrum of the CMB gravitational lensing potential. For this, we use the latest and most precise data combination from the \textit{Planck} 2018 lensing reconstruction and the Data Release 6 (DR6) of the Atacama Cosmology Telescope (ACT) \citep{ACT:2023kun,ACT:2025fju}. This combined likelihood provides powerful complementary constraints on the geometry of the late Universe and the growth of structures.

\subsection{Type Ia Supernovae}

Type Ia supernovae are powerful cosmological probes, acting as standardizable candles that allow us to measure the expansion history of the universe through the luminosity distance-redshift relation. For this analysis, we use the recent Dark Energy Survey Year 5 (DESY5) data release \citep{DES:2024jxu}.

The DESY5 dataset is notable for its size and uniformity. It consists of a homogeneously calibrated sample of 1635 photometrically-classified SNe Ia with redshifts in the range $0.1 < z < 1.13$, observed by a single survey. This main sample is complemented by a set of 194 low-redshift ($z < 0.1$) SNe Ia from external surveys to anchor the Hubble diagram at low redshift. The light curves in the DESY5 analysis are processed using the SALT3 model \citep{Kenworthy:2021azy}.

The cosmological constraints from SNe Ia are derived by comparing the observed distance modulus, $\mu_{\text{obs}}$, with the theoretical prediction, $\mu_{\text{th}}$. The theoretical distance modulus for a given set of cosmological parameters $\theta$ is:
\begin{equation}
    \mu_{\text{th}}(z; \theta) = 5 \log_{10}\left(\frac{d_L(z; \theta)}{\text{Mpc}}\right) + 25,
\end{equation}
where $d_L(z; \theta)$ is the luminosity distance, which depends on the underlying cosmological model. The likelihood for the SN Ia sample is given by the $\chi^2$ function:
\begin{equation}
    \chi^2_{\text{SNe}} = \Delta\boldsymbol{\mu}^{T} \cdot \mathbf{C}^{-1} \cdot \Delta\boldsymbol{\mu}.
\end{equation}
Here, $\Delta\boldsymbol{\mu} = \boldsymbol{\mu}_{\text{obs}} - \boldsymbol{\mu}_{\text{th}}$ is the vector of residuals between the observed and theoretical distance moduli for all supernovae. The term $\mathbf{C}$ is the total covariance matrix, which includes both statistical uncertainties from the light-curve fitting and systematic uncertainties associated with calibration, astrophysical effects, and modeling choices. The absolute magnitude of SNe Ia, $M$, which is degenerate with the Hubble constant, is treated as a nuisance parameter and analytically marginalized over within the likelihood calculation.

\subsection{Baryon Acoustic Oscillations}

\begin{table}
\centering
\setlength{\tabcolsep}{12pt}
\renewcommand{\arraystretch}{1.2}
\caption{DESI DR2 BAO distance measurements.}
\label{tab:desi_bao_data_compact}
\begin{tabular}{l |c |l}
\toprule
Tracer & $z_{\text{eff}}$ & Measurement \\
\toprule
BGS    & 0.295            & $D_V/r_d = 7.942 \pm 0.075$ \\
\midrule
\multirow{2}{*}{LRG1} & \multirow{2}{*}{0.510} & $D_M/r_d = 13.59 \pm 0.17$ \\
                      &                        & $D_H/r_d = 21.86 \pm 0.43$ \\
\midrule
\multirow{2}{*}{LRG2} & \multirow{2}{*}{0.706} & $D_M/r_d = 17.35 \pm 0.18$ \\
                      &                        & $D_H/r_d = 19.46 \pm 0.33$ \\
\midrule
\multirow{2}{*}{LRG+ELG} & \multirow{2}{*}{0.934} & $D_M/r_d = 21.58 \pm 0.15$ \\
                         &                        & $D_H/r_d = 17.64 \pm 0.19$ \\
\midrule
\multirow{2}{*}{ELG} & \multirow{2}{*}{0.934} & $D_M/r_d = 27.60 \pm 0.32$ \\
                     &                        & $D_H/r_d = 14.18 \pm 0.22$ \\
\midrule
\multirow{2}{*}{QSO} & \multirow{2}{*}{1.484} & $D_M/r_d = 30.51 \pm 0.76$ \\
                     &                        & $D_H/r_d = 12.82 \pm 0.52$ \\
\midrule
\multirow{2}{*}{Ly$\alpha$ Forest} & \multirow{2}{*}{2.330} & $D_M/r_d = 38.99 \pm 0.53$ \\
                     &                        & $D_H/r_d = 8.632 \pm 0.101$ \\
\bottomrule
\end{tabular}
\end{table}

Baryon Acoustic Oscillations provide a powerful standard ruler to map the expansion history of the Universe. The characteristic scale of these oscillations, imprinted in the early Universe and visible today in the clustering of matter, is the sound horizon at the drag epoch, $r_d$. By measuring the apparent size of this scale at different redshifts, we can constrain the transverse comoving distance, $D_M(z)$, and the Hubble parameter, $H(z)$.

In this work, we use the final consensus BAO measurements from the DESI DR2 \citep{DESI:2025zgx}. This dataset represents a significant leap in precision, providing the most accurate BAO distance measurements to date. The measurements are derived from the two-point correlation function of several distinct classes of spectroscopic targets (tracers), which together cover a wide redshift range: the Bright Galaxy Sample (BGS), Luminous Red Galaxies (LRGs), Emission Line Galaxies (ELGs), Quasars (QSOs), and the Lyman-$\alpha$ (Ly$\alpha$) forest absorption in the spectra of high-redshift quasars.

The DESI collaboration provides the distilled cosmological information as a set of measurements of $D_M(z)/r_d$ and $D_H(z)/r_d$ (where $D_H(z) = c/H(z)$) at seven effective redshifts, from $z=0.295$ to $z=2.33$. For the lowest redshift BGS sample, only the isotropic volume-averaged distance, $D_V(z)/r_d$, is constrained, where $D_V(z) \equiv [cz D_M(z)^2/H(z)]^{1/3}$.

These measurements are encapsulated in a data vector, $\mathbf{d}_{\text{obs}}$, and their statistical and systematic uncertainties (including correlations between all points) are described by a total covariance matrix, $\mathbf{C}$. The likelihood is then given by the $\chi^2$ function:
\begin{equation}
    \chi^2_{\text{BAO}} = \left(\mathbf{d}_{\text{obs}} - \mathbf{d}_{\text{th}}(\theta)\right)^T \cdot \mathbf{C}^{-1} \cdot \left(\mathbf{d}_{\text{obs}} - \mathbf{d}_{\text{th}}(\theta)\right).
\end{equation}
Here, $\mathbf{d}_{\text{th}}(\theta)$ is the vector of theoretical predictions for the distance ratios, calculated for a given set of cosmological parameters $\theta$. The specific data points used in our analysis are summarized in Table \ref{tab:desi_bao_data_compact}.

\section{Results}\label{sec:results}

\subsection{Analysis Setup}

We perform a Bayesian statistical analysis to infer the cosmological parameters for three distinct models: the standard flat $\Lambda$CDM model, the CPL ($w_0w_a$CDM) model, and our  Interaction Model. To this end, we use the MCMC sampler implemented in the \texttt{Cobaya} software \citep{Torrado:2020dgo}, coupled with a modified version of the \texttt{CLASS} Einstein-Boltzmann solver \citep{Blas:2011rf} that incorporates our theoretical formalism.
For our main analysis, we use a comprehensive combination of the latest cosmological data, which we refer to as our baseline dataset. This data is the same as describe in Section \ref{sec:data}.

In the flat $\Lambda$CDM model, we vary the six standard cosmological parameters. For the extended models (CPL and Interaction), we vary these six plus two additional parameters describing the dark energy sector, $w_0$ and $w_a$. It is important to emphasize that, although the same parameters are employed in both the dynamical and interacting approaches, the physical meaning of $w_0$ and $w_a$ differs in each case. In the dynamical description of dark energy, these parameters characterize the dark energy equation of state, whereas in the interacting scenario they encode the intensity and form of the dark sector interaction. In both frameworks, the $\Lambda$CDM limit is recovered when $w_0 = -1$ and $w_a = 0$, corresponding to a non-interacting cosmological constant.

Finally, the sum of neutrino masses is held fixed at its minimal value allowed by oscillation experiments, $\sum m_\nu = 0.06$~eV. We use wide, uninformative flat priors for all varied parameters, as summarized in Table~\ref{tab:priors}. The convergence of the MCMC chains is assessed using the Gelman-Rubin statistic, for which we require $R-1 < 0.01$.

\begin{table}
\centering
\setlength{\tabcolsep}{20 pt}
\renewcommand{\arraystretch}{1.2}
\caption{Parameters varied in our MCMC analysis and their corresponding prior distributions. The top block lists the six base parameters for the $\Lambda$CDM model. The bottom block lists the additional parameters for the extended models. $U[a, b]$ denotes a uniform (flat) prior between $a$ and $b$.}
\label{tab:priors}
\begin{tabular}{l c}
\toprule
Parameter & Prior \\
\midrule
\multicolumn{2}{c}{\textbf{Base Parameters}} \\
$\Omega_b h^2$ & $U[0.005, 0.1]$ \\
$\Omega_c h^2$ & $U[0.001, 0.99]$ \\
$100\theta_{s}$ & $U[0.5, 10]$ \\
$\ln(10^{10} A_s)$ & $U[1.61, 3.91]$ \\
$n_s$ & $U[0.8, 1.2]$ \\
$\tau_{\text{reio}}$ & $U[0.01, 0.8]$ \\
\midrule
\multicolumn{2}{c}{\textbf{Extended Parameters}} \\
$w_0$ & $U[-3, 1]$ \\
$w_a$ & $U[-3, 2]$ \\
\bottomrule
\end{tabular}
\end{table}

\subsection{Parameter Constraints and Model Degeneracies}

\begin{table*}[t]
    \centering
    \setlength{\tabcolsep}{14pt}
\renewcommand{\arraystretch}{1.5}
    \caption{Constraints for the $\Lambda$CDM, CPL, and Interaction models from our baseline dataset combination (CMB+BAO+SNe). The errors correspond to the 68\% and 95\% C.L.\ intervals.}
    \label{tab:param_constraints}
    \begin{tabular}{l c c c}
        \toprule
        Parameter & $\Lambda$CDM & CPL Model & Interaction Model \\
        \midrule
        $w_0$ & $\equiv -1$ & $-0.76^{+0.20+0.22}_{-0.20-0.22}$ & $-0.74^{+0.19+0.22}_{-0.20-0.25}$ \\
        $w_a$ & $\equiv 0$ & $-0.92^{+0.76+0.85}_{-0.69-0.81}$ & $-0.85^{+0.74+0.84}_{-0.78-0.97}$ \\
        \midrule
        $H_0$ [km/s/Mpc] & $67.92^{+0.96+1.04}_{-0.84-0.99}$ & $67.0^{+1.6+1.8}_{-2.2-2.4}$ & $66.5^{+1.9+2.2}_{-1.7-1.9}$ \\
        $\Omega_m$ & $0.306^{+0.012+0.013}_{-0.011-0.013}$ & $0.317^{+0.023+0.024}_{-0.018-0.019}$ & $0.49^{+0.13+0.16}_{-0.15-0.18}$ \\
        $\sigma_8$ & $0.813^{+0.014+0.016}_{-0.019-0.021}$ & $0.809^{+0.024+0.024}_{-0.030-0.036}$ & $0.53^{+0.19+0.27}_{-0.11-0.11}$ \\
        \bottomrule
    \end{tabular}
\end{table*}

\begin{figure}
    \centering
    \includegraphics[width=\linewidth]{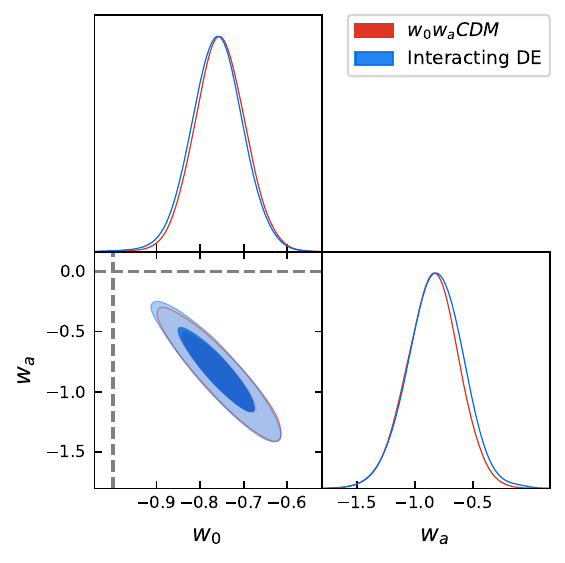}
    \caption{The 68\% and 95\% C.L.\ posterior contours for the $(w_0, w_a)$ parameters. The CPL model is shown in red, and the Interaction model in blue. The black dashed lines indicate the $\Lambda$CDM model, which is excluded at high significance.}
    \label{fig:w0wa}
\end{figure}

\begin{figure}
    \centering
    \includegraphics[width=\linewidth]{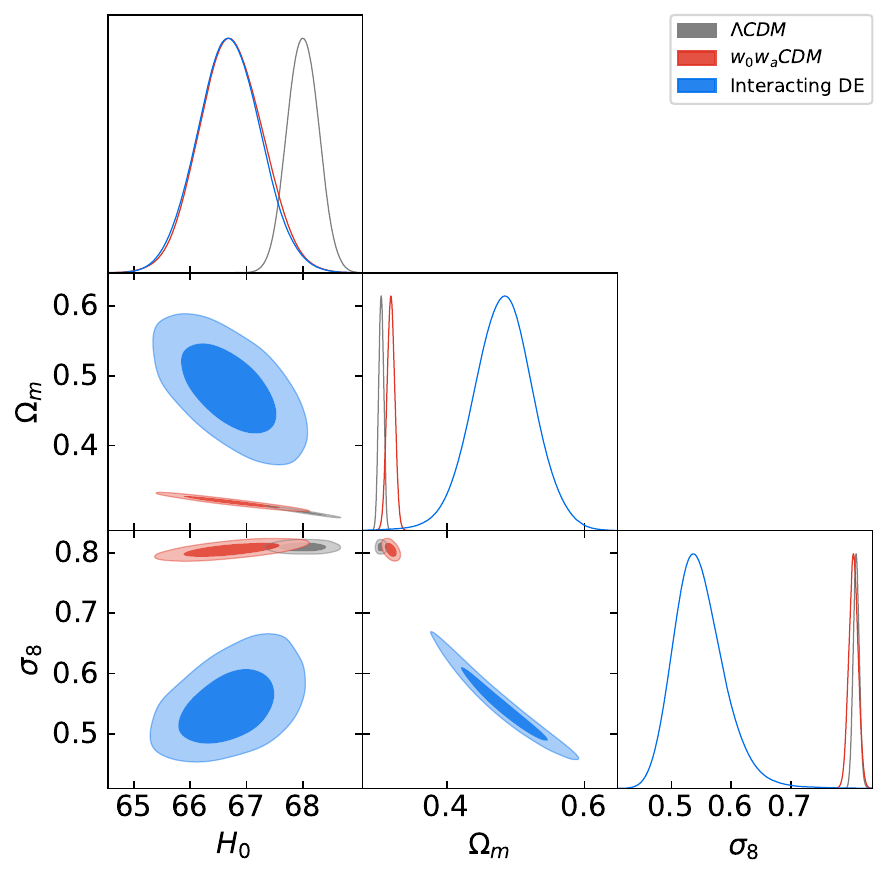}
    \caption{The 68\% and 95\% C.L.\ posterior contours for $H_0$, $\Omega_m$, and $\sigma_8$. The results for $\Lambda$CDM, CPL, and the Interaction model are shown.}
    \label{fig:triplot}
\end{figure}

\begin{figure}
\centering
\includegraphics[width=\linewidth]{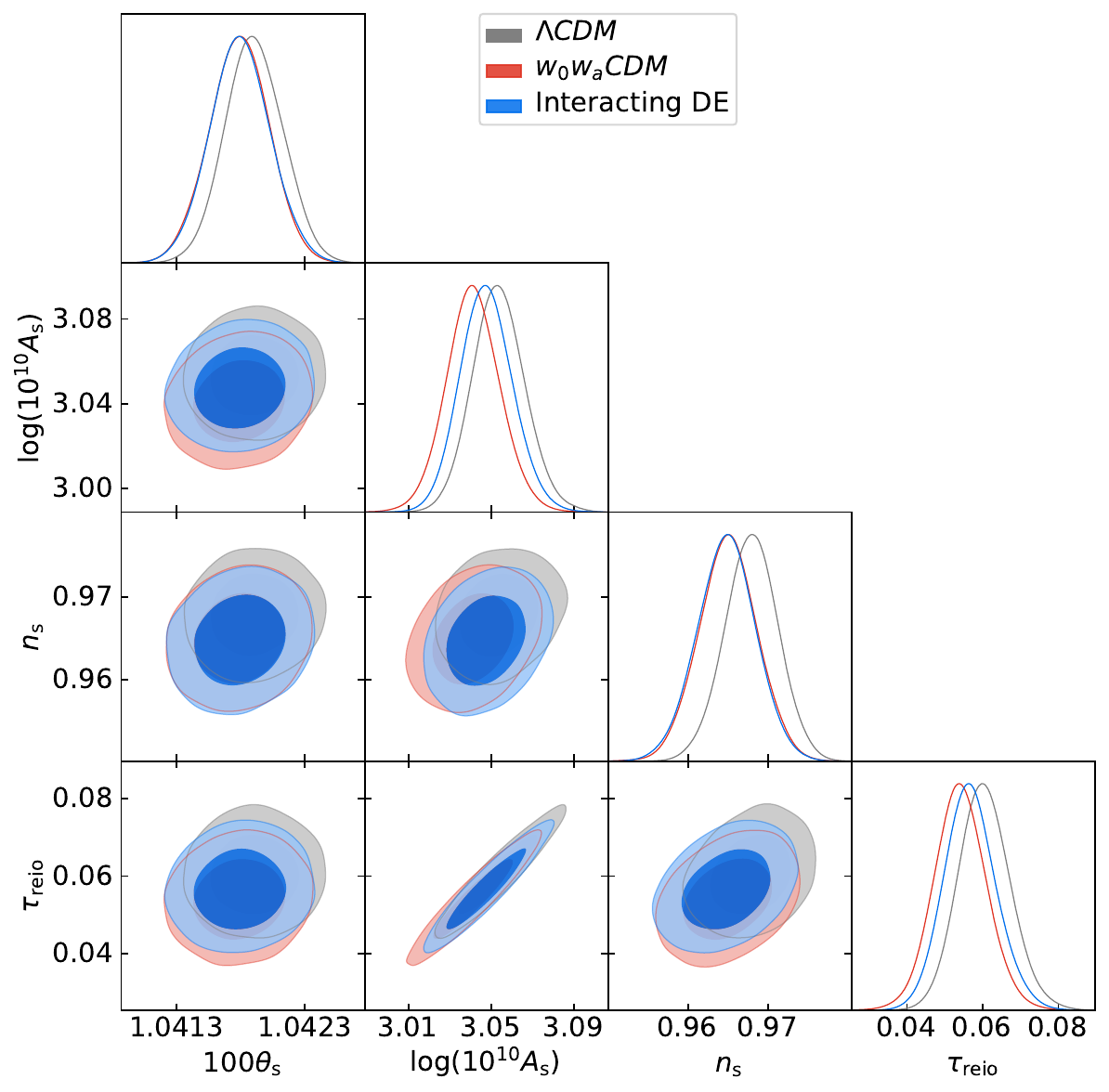}
\caption{The 68\% and 95\% C.L.\ posterior contours for the early-Universe parameters: the angular acoustic scale ($\theta_s$), scalar amplitude ($\log(10^{10}A_s)$), spectral index ($n_s$), and optical depth to reionization ($\tau_{\mathrm{reio}}$).}
\label{fig:triplot-eu}
\end{figure}

We present the main cosmological constraints from our MCMC analysis. Our results are derived from the baseline dataset combination (CMB+BAO+SNe) described in Section~\ref{sec:data}. The marginalized constraints for the key parameters of the $\Lambda$CDM, CPL, and Interaction models are summarized in Table~\ref{tab:param_constraints}. The full posterior distributions and the correlations between the most relevant parameters are shown in Figs.~\ref{fig:w0wa}-\ref{fig:triplot-eu}. In these and subsequent figures and tables, we omit bars and tildes in the parameter labels for simplicity; however, constraints for the Interaction model refer to tilde quantities, and those for the dynamical model to barred quantities.

The primary result for the extended models are the constraints on the dark energy parameters, which are detailed in Table~\ref{tab:param_constraints} and shown in Fig.~\ref{fig:w0wa}. Remarkably, our Interaction model and the CPL model yield nearly identical constraints. The posterior contours for the $(w_0, w_a)$ pair for both models overlap almost perfectly, highlighting a strong degeneracy at the level of the background parameterization. For both models, the $\Lambda$CDM fixed point of $(w_0, w_a) = (-1, 0)$ is excluded at more than 2$\sigma$ confidence, indicating a statistical preference of the baseline dataset for models beyond $\Lambda$CDM.

However, a crucial finding of this work emerges when examining other cosmological parameters (Fig.~\ref{fig:triplot}). While the posteriors for the Hubble constant relative to the Interaction and CPL models are basically identical, and slightly shifted to lower values with respect to $\Lambda$CDM, the Interaction model yields significantly different values for the present-day matter density, $\Omega_m$, and the amplitude of matter fluctuations, $\sigma_8$. As seen in Table~\ref{tab:param_constraints}, the Interaction model prefers a much higher $\Omega_m$ and a lower $\sigma_8$ compared to both $\Lambda$CDM and the CPL model. This result breaks the degeneracy between the two extended models, demonstrating that although they can be parameterized in a similar way, their distinct physical natures lead to different impacts on the cosmic matter content and its clustering. In this sense, an observable that constrains only the matter sector, rather than the Hubble rate or the gravitational potential, could allow us to distinguish between the two competing descriptions of the dark sector: dynamical dark energy and an interacting vacuum \citep{Barbosa:2024ppn}.

Regarding the tension in the Hubble constant between the value inferred from the CMB under the $\Lambda$CDM assumption (e.g., $H_0 = 67.4 \pm 0.5$ km/s/Mpc from \textit{Planck} \citep{Planck:2018vyg}) and the value measured locally (e.g., $H_0 = 73.0 \pm 1.0$ km/s/Mpc from the SH0ES program \citep{Riess:2021jrx}), all three models, including the Interaction model, favor a low $H_0$ consistent with the \textit{Planck} results.  
We therefore conclude that the proposed interaction mechanism does not alleviate the $H_0$ tension, in agreement with expectations for this class of interacting dark energy models~\citep{Lee:2022cyh}.

\begin{figure}
\centering
\includegraphics[width=\linewidth]{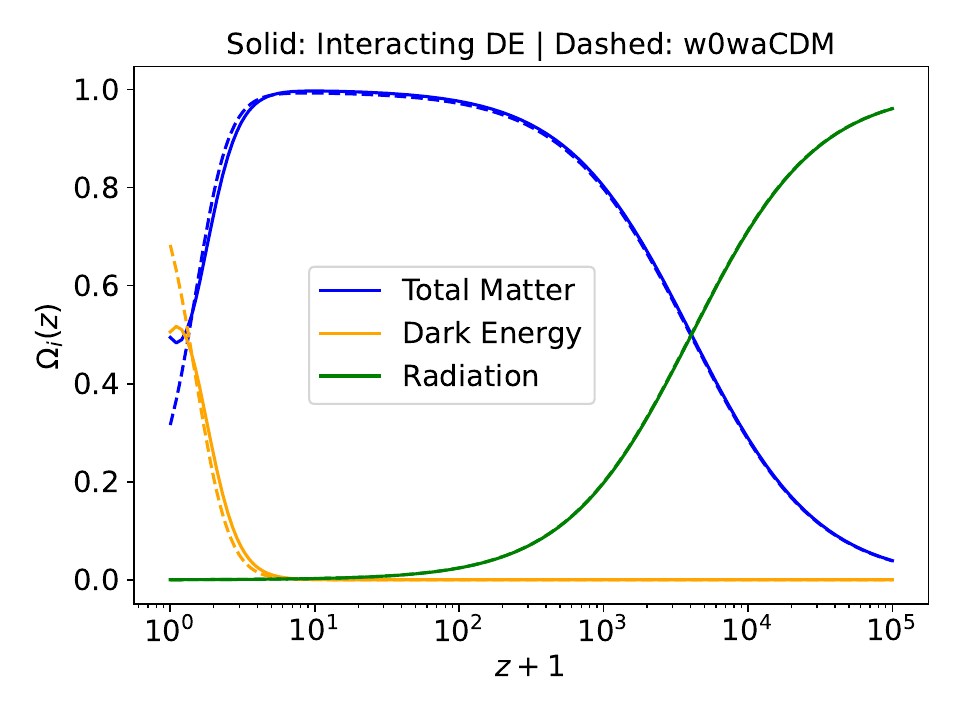}
\caption{Evolutions of the CPL and IDE best-fit models.}
\label{fig:omegas}
\end{figure}

On the other hand, the results for the early-Universe parameters shown in Fig.~\ref{fig:triplot-eu} indicate that the different models yield nearly identical constraints. This confirms that the impact of the interacting dark energy model is confined mainly to low redshifts ($z \lesssim 5$), leaving the physics of the early Universe essentially unchanged. This conclusion is further supported by Fig.~\ref{fig:omegas}, which shows that the evolution histories of the CPL and IDE models are indistinguishable at early times.

\begin{figure}
\centering
\includegraphics[width=\columnwidth]{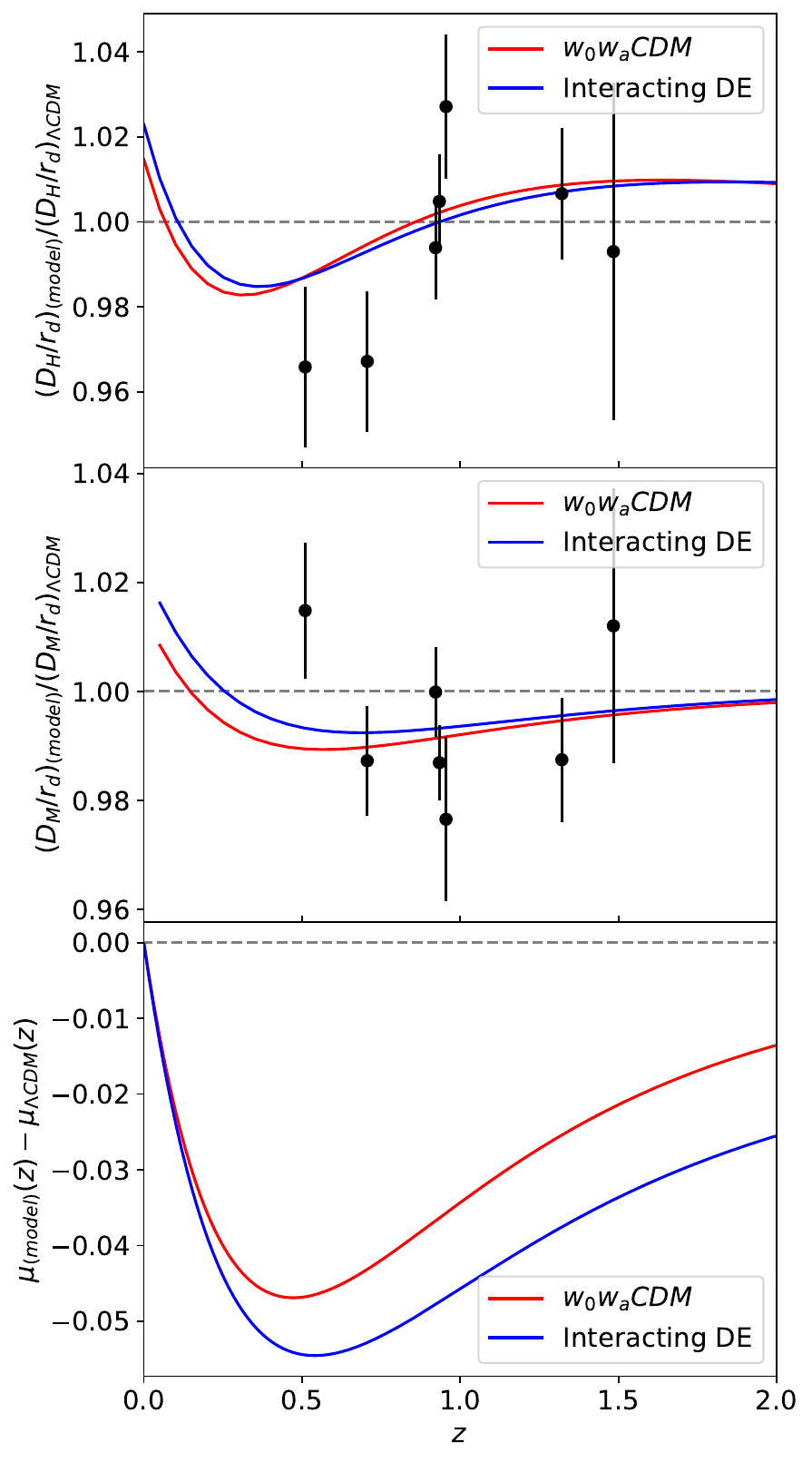}
\caption{Comparison of the background expansion history for the CPL and Interaction models, relative to the best-fit $\Lambda$CDM model.
The points in the top and center panels are from \cite[][Table IV]{DESI:2025zgx}.}
\label{fig:expansion_probes}
\end{figure}

This stark difference in the matter sector parameters is particularly striking given that the models are degenerate in their background expansion history for construction. We demonstrate this visually in Figure~\ref{fig:expansion_probes}. The figure shows the key geometric probes: the distance modulus $\mu(z)$, the Hubble distance $D_H(z)/r_d$, and the transverse comoving distance $D_M(z)/r_d$, as ratios relative to the best-fit $\Lambda$CDM model. For both the CPL and Interaction models, the deviations from $\Lambda$CDM are minimal, at the level of a few percent at most across the entire redshift range probed by the data.
The observed difference arises from fluctuations in parameter estimation and, more importantly, from the fact that CMB data partially breaks the background-level degeneracy between the two models, due to integrated effects. This reinforces the need for structure growth probes to fully disentangle them.

\subsubsection{Physical Interpretation of the Background Results}

\begin{figure}
\centering
\includegraphics[width=\linewidth]{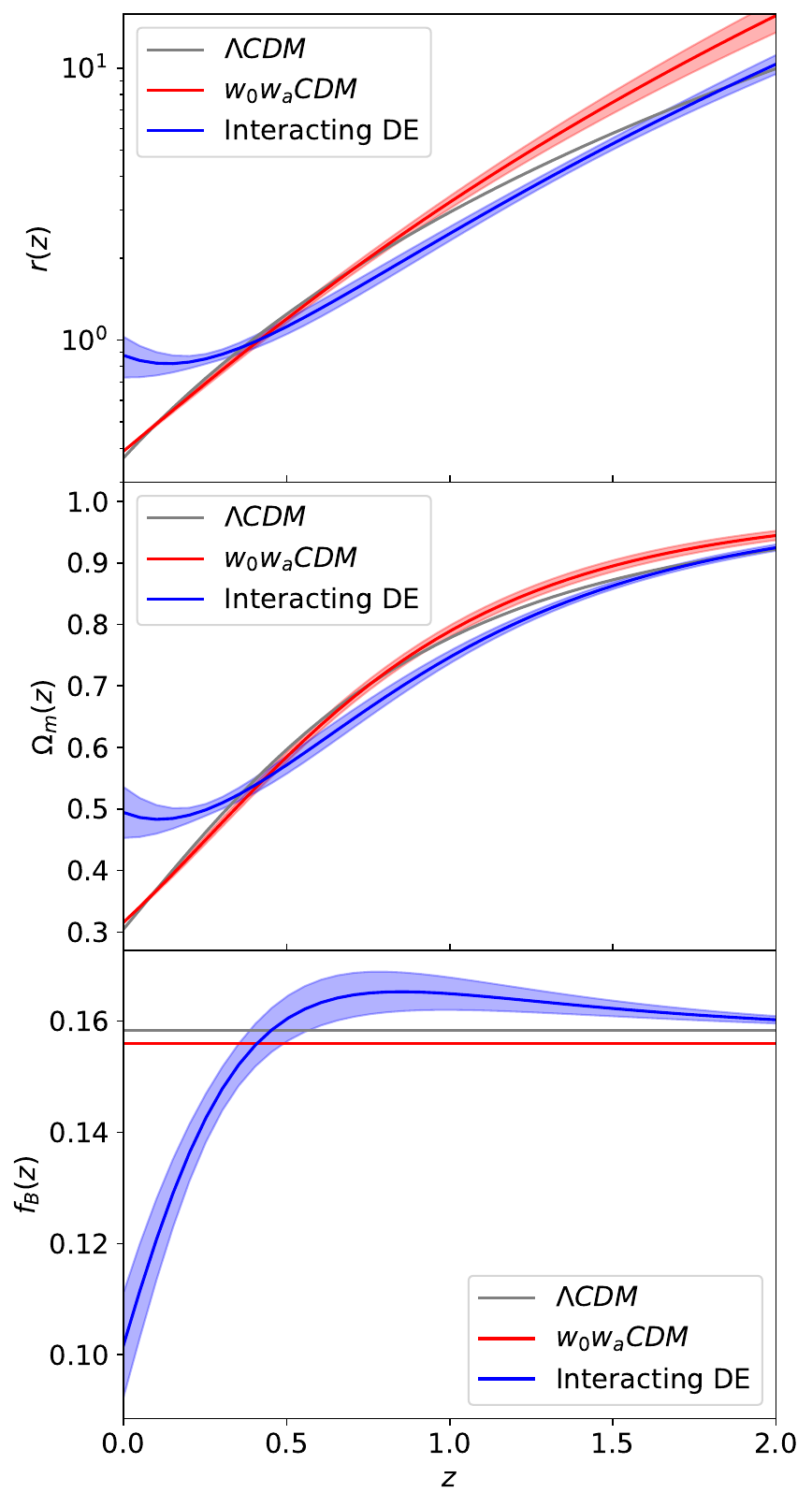}
\caption{Evolution of the dark sector energy density ratio $r = \rho_c/\rho_x$ (top), the matter density parameter  $\Omega_m$ (middle) and the baryon fraction parameter $f_B = \rho_b /(\rho_b + \rho_c)$ (bottom).}
\label{fig:r}
\end{figure}

To understand the origin of the discrepancy in $\Omega_m$ and $\sigma_8$, we investigated the background-level evolution of the models. For the plots in this subsection, the central curves are derived from the best-fit parameters of each model, while the vertical error bars or shaded regions represent the 68\% C.L. uncertainty propagated from the MCMC chains for $w_0$ and $w_a$.

We began by analyzing the evolution of the dark sector energy density ratio, $r(z) = \rho_c/\rho_x$, shown in Figure~\ref{fig:r} (top panel). While the CPL model shows a monotonic decrease, the Interaction model's ratio exhibits a distinct change in behavior for $z < 0.5$, beginning to grow towards the present day. This hints at a change in the dark sector dynamics.

\begin{figure}
\centering
\includegraphics[width=\linewidth]{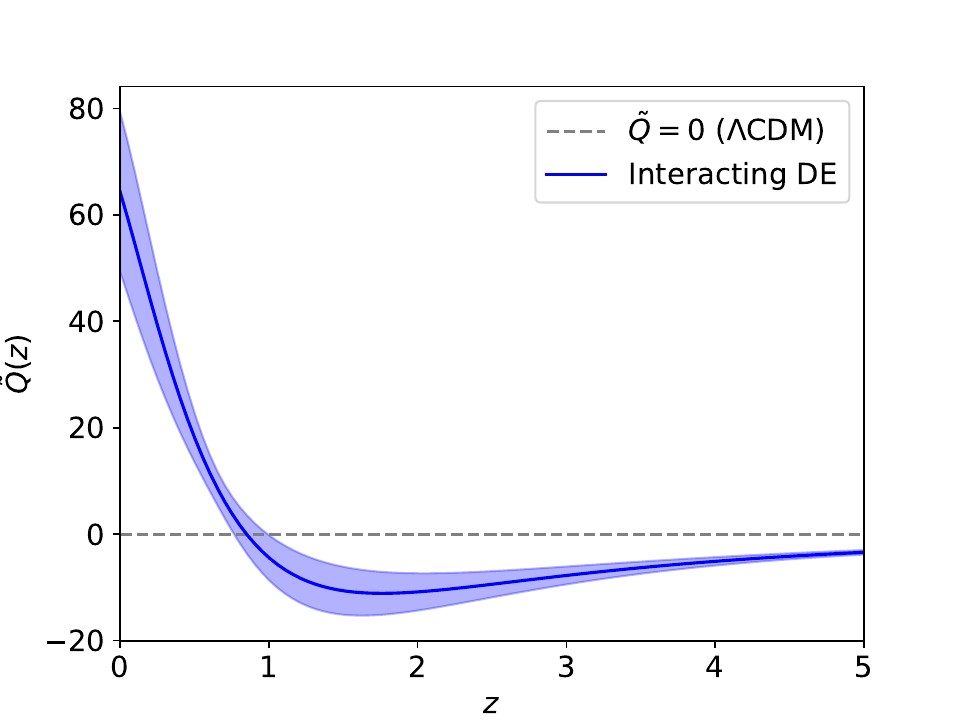}
\caption{Evolution of the interaction term $\tilde{Q}(z)$.}
\label{fig:tilde_q}
\end{figure}

The key physical component of the Interaction model is the interaction term, $\tilde{Q}(z)$, which dictates the energy exchange. Figure~\ref{fig:tilde_q} reveals that $\tilde{Q}(z)$ undergoes a sign change at a redshift of $z \approx 0.8$ \citep{Li:2025owk}.
This sign-change event has profound physical implications. For $z > 0.8$, the interaction term is negative, corresponding to a scenario where dark matter decays into dark energy. However, for $z < 0.8$, the interaction term becomes positive, reversing the direction of energy flow. In this late-time epoch, dark energy begins to decay into dark matter. It is worth mentioning that the sign change of the interaction is associated with the $-1$ crossing of the dynamical dark energy equation of state. In this sense, one may argue that the interacting approach is more physical, as it naturally avoids the $-1$ crossing. See~\citet{Ye:2024ywg,Wolf:2025jed} for models that realize phantom crossing.

This late-time energy injection into the dark matter sector directly explains the MCMC results. The continuous creation of dark matter at low redshift leads to a significantly higher value for its present-day energy density, $\Omega_{m,0}$ (Fig.~\ref{fig:r}, mid panel). This mechanism also causes the cosmic baryon fraction, $f_b(z) = \rho_b(z) / (\rho_b(z) + \rho_c(z))$, to decrease at late times (Fig.~\ref{fig:r}, bottom panel).

\subsubsection{Signatures in Large-Scale Structure}

\begin{figure}
\centering
\includegraphics[width=\linewidth]{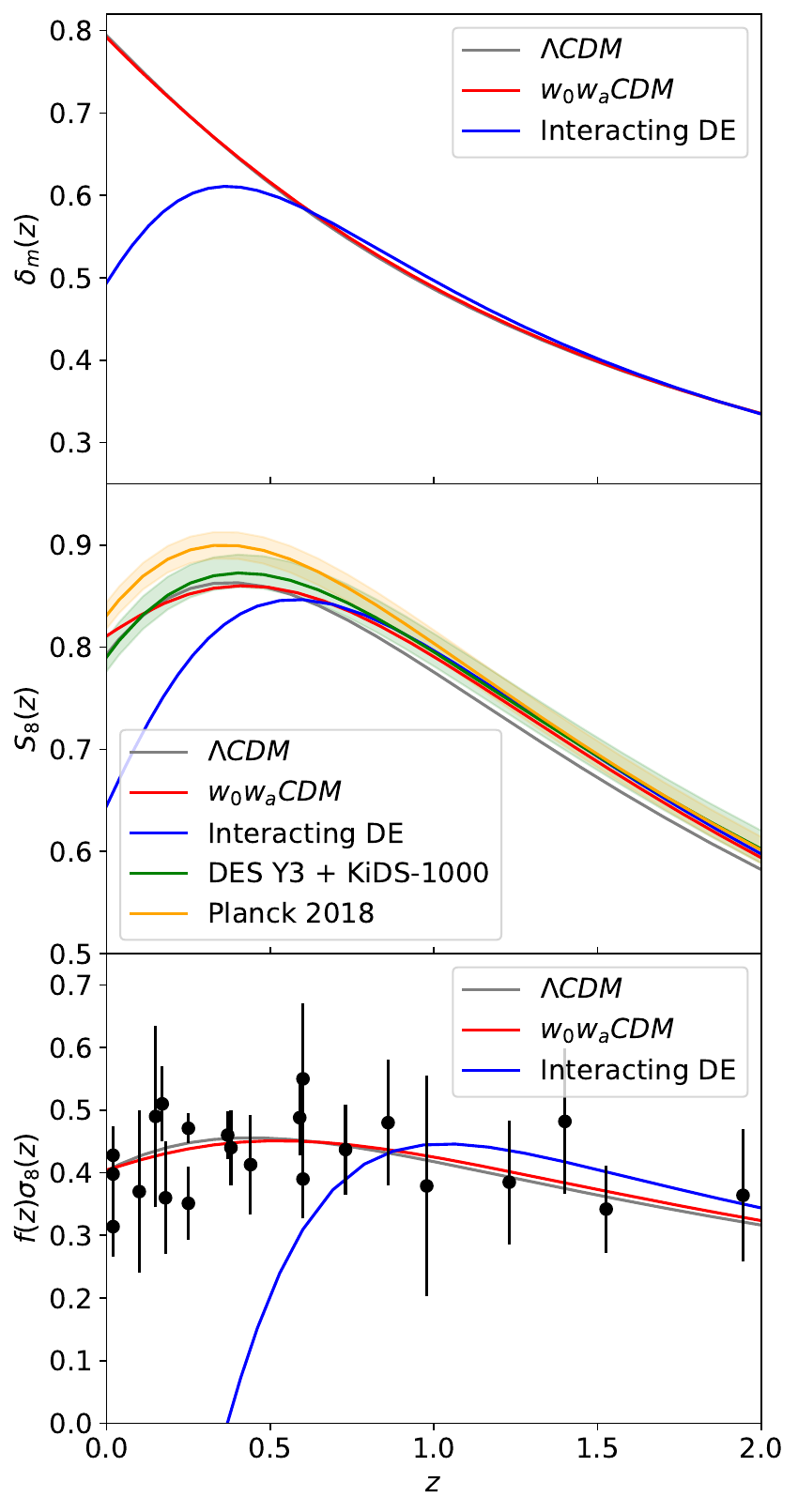}
\caption{Evolution of the linear total matter perturbation $\delta_t(z)$, normalized to the matter-dominated epoch (top), the matter clustering amplitude $S_8(z)$ (middle), and the growth rate $f\sigma_8(z)$ (bottom).
The shaded band in the middle panel shows the 68\% C.L.\ constraints from Planck 2018 \citep{Planck:2018vyg} and the joint cosmic shear analysis of DES Y3 and KiDS-1000 \citep{Kilo-DegreeSurvey:2023gfr}.
The bottom panel includes measurements from various  surveys \citep{Akarsu:2025ijk}.}
\label{fig:delta_t}
\end{figure}

The dramatic modification of the background matter density in the Interaction model leads to distinct and testable predictions for the growth of large-scale structure. We computed the evolution of linear matter perturbations using the best-fit parameters for each model within \texttt{CLASS}. Figure~\ref{fig:delta_t} (top panel) shows the evolution of the total matter density contrast (baryons + dark matter), $\delta_m(z)$. The figure clearly shows that the anomalous background evolution in the Interaction model leads to a visible suppression of growth at late times ($z < 0.8$) when compared to both $\Lambda$CDM and the CPL model.

This suppression has a clear impact on the structure growth amplitude, $S_8(z) = \sigma_8(z)\sqrt{\Omega_m(z)/0.3}$, shown in Figure~\ref{fig:delta_t} (middle panel). The CPL model remains in tension with the joint cosmic shear constraints from DES Y3 and KiDS-1000~\citep{Kilo-DegreeSurvey:2023gfr}.  
Late-Universe probes, such as weak lensing surveys from DES and KiDS, typically find lower values ($S_8 \approx 0.76$--0.79) than those inferred from the \textit{Planck} CMB data ($S_8 \approx 0.83$).  
The Interaction model, in contrast, follows the CPL behavior at higher redshifts but, as the interaction reverses sign around $z \approx 0.8$, $S_8(z)$ decreases sharply, crossing the cosmic shear data band and evolving toward a lower present-day value.  
It is worth noting that DES Y3 and KiDS-1000 have a median redshift of 0.6--0.7, where the Interaction model remains consistent with the observational constraints.  
Overall, the Interaction model suggests the potential to accommodate the lower $S_8$ values reported by weak lensing surveys, although these data were not included in our fit.

Figure~\ref{fig:delta_t} (bottom panel) shows the growth rate parameter combination $f\sigma_8(z)$, where $f(z)$ is the linear growth rate and $\sigma_8(z)$ is the RMS amplitude of matter fluctuations at a given redshift. While the $\Lambda$CDM and CPL models predict a slowly varying evolution, our Interaction model predicts a sharp drop a $z < 0.8$, coinciding with the epoch where dark energy begins to decay into dark matter.
This  behavior was previously found in models that feature non-adiabatic dark energy perturbations \citep{Zimdahl:2019pqg}.
This unique prediction makes the Interaction model highly falsifiable. Indeed, the figure shows that the predicted sharp decline is in significant tension with current growth rate data from various surveys, as compiled in \cite[][Table IV]{Akarsu:2025ijk}.

This suggests that the model cannot simultaneously fit the CMB and background expansion data (BAO and SNe) together with the observed history of structure formation.
However, our statistical inference did not include the $f\sigma_8$ measurements in the fit, and a better agreement might be obtained in a more comprehensive analysis.
Such an analysis would require a careful examination of the fiducial cosmologies used to derive the data in \cite[][Table IV]{Akarsu:2025ijk}, which we leave for future work.

\subsection{Model Comparison}

While the previous section highlighted the distinct physical predictions of the CPL and Interaction models, their nearly identical constraints on the $(w_0, w_a)$ parameters necessitate a formal statistical comparison. To evaluate the performance of the extended models against the standard $\Lambda$CDM, we use information criteria, which balance the goodness-of-fit against model complexity.

The goodness-of-fit for each model is given by the minimum chi-squared, $\chi^2_{\text{min}}$, obtained from the MCMC analysis. We find that both extended models provide a better fit to the data than $\Lambda$CDM, as shown in Table~\ref{tab:model_comparison}. These are significant improvements, but they come at the cost of two additional free parameters. Although the dynamical and interacting approaches yield nearly identical CMB spectra, given their similarity at recombination, the differences observed in the statistical analysis are justified by the contributions from the integrated Sachs–Wolfe effect and CMB lensing.

To properly assess whether this improvement is statistically justified, we compute two information criteria. The first is the Akaike Information Criterion (AIC), defined as:
\begin{equation}
    \text{AIC} = \chi^2_{\text{min}} + 2k,
\end{equation}
where $k$ is the number of free cosmological parameters.

The second criterion is the Deviance Information Criterion (DIC) \citep{Spiegelhalter2002}. The DIC is particularly well-suited for Bayesian model comparison using MCMC outputs and is designed to handle complex, correlated datasets, such as the \textit{Planck} CMB likelihoods used in this analysis \citep{linddle2007}. It uses a data-driven penalty based on the effective number of parameters, $p_D = \overline{\chi^2} - \chi^2_{\text{min}}$, where $\overline{\chi^2}$ is the mean $\chi^2$ averaged over the posterior. The DIC is then defined as:
\begin{equation}
    \text{DIC} = \chi^2_{\text{min}} + 2p_D.
\end{equation}

We compare models using the difference in these criteria with respect to $\Lambda$CDM, $\Delta\text{IC} = \text{IC}_{\text{model}} - \text{IC}_{\Lambda\text{CDM}}$. A negative value of $\Delta\text{IC}$ indicates a preference for the extended model.
The results of our model comparison are summarized in Table~\ref{tab:model_comparison}. Both the CPL and Interaction models are strongly preferred over $\Lambda$CDM according to the AIC. This indicates that the inclusion of two extra parameters is well justified by the significant improvement in the fit to the data.
The DIC, which more robustly handles the full dataset, provides an even stronger conclusion. Both the CPL model  and the Interaction model are decisively preferred over $\Lambda$CDM. This confirms that the inclusion of two new parameters is strongly favored by the data.

In summary, while the data shows a strong statistical preference for a deviation from the standard $\Lambda$CDM model, information criteria cannot unambiguously distinguish between a dynamical dark energy fluid and the proposed dark sector interaction. Between the two new models, the CPL model is marginally preferred, but the statistical difference is too small to draw a firm conclusion. This reinforces that the two scenarios remain largely degenerate from a statistical standpoint, despite their different physical implications.

\begin{table}
    \centering
    \setlength{\tabcolsep}{5pt} 
    \renewcommand{\arraystretch}{1.4}
    \caption{Model comparison statistics for the three models, using $\Lambda$CDM as the reference. }
    \label{tab:model_comparison}
    \begin{tabular}{l c c c c c}
        \toprule
        Model & $k$ & $\chi^2_{\text{min}}$ & $\Delta\chi^2$ & $\Delta\text{AIC}$ & $\Delta\text{DIC}$ \\
        \midrule
        $\Lambda$CDM & 6 & 12660.9 & -- & -- & -- \\
        CPL Model & 8 & 12646.4 & -14.5 & -10.5 & -11.1 \\
        Interaction Model & 8 & 12649.52 & -11.4 & -7.4 & -10.0 \\
        \bottomrule
    \end{tabular}
\end{table}

\section{Conclusions}\label{sec:conclusion}

In this work, we have investigated a specific model of a dark sector interaction, where dark energy and dark matter exchange energy throughout cosmic history. The model is constructed to be degenerate at the background level with the well-known CPL parameterization, allowing for a direct test of how distinct physical mechanisms can be constrained by cosmological data, even when they predict the same expansion history. We performed a comprehensive MCMC analysis, constraining the model parameters with a combination of the latest cosmological datasets, including CMB data from \textit{Planck} and ACT, BAO measurements from DESI DR2, and Type Ia Supernovae from the DESY5 survey.

Our key findings can be summarized as follows:

\begin{enumerate}
    \item \textbf{A Better Fit than $\Lambda$CDM:} We found that both the Interaction model and its degenerate CPL counterpart provide a significantly better fit to the combined dataset than the standard $\Lambda$CDM model. Information criteria tests support this conclusion, with the AIC and DIC showing decisive preference for the extended models.

    \item \textbf{Breaking the Degeneracy:} While the two extended models are statistically indistinguishable based on their expansion history parameters $(w_0, w_a)$, we demonstrated that the degeneracy is broken when considering the matter sector. The Interaction model predicts a significantly higher present-day matter density ($\Omega_m$) and a lower amplitude of matter fluctuations ($\sigma_8$) compared to both $\Lambda$CDM and the CPL model.

    \item \textbf{A Physical Mechanism with Testable Consequences:} We identified the physical origin of this broken degeneracy: a sign change in the interaction term $\tilde{Q}(z)$ at low redshift ($z \approx 0.8$), which reverses the energy flow from dark matter decaying into dark energy to the opposite process. This transition is directly tied to the $w=-1$ crossing in the CPL parameterization; in this sense, the interacting approach may be viewed as more physical, since it avoids the problematic crossing by construction  \citep{Guedezounme:2025wav,Wang:2025znm}. The late-time injection of dark matter has profound and testable implications.

    \item \textbf{Alleviation of the $S_8$ Tension:} The combination of a higher $\Omega_m$ and a lower $\sigma_8$ leads the Interaction model to predict a lower value of $S_8$ on the redshift range probed by weak lensing data, although we emphasize that these data were not included in our fit.

    \item \textbf{A New Tension with Growth of Structure:} The same mechanism that alleviates the $S_8$ tension leads to a unique prediction for the evolution of the growth rate, $f\sigma_8(z)$. The model predicts a sharp suppression of growth at $z < 0.8$, which is in  disagreement with  measurements of the growth rate.

\end{enumerate}

In conclusion, the interacting dark energy model analyzed in this work provides an illustrative case study.
It offers a better fit to the background and CMB data than the standard $\Lambda$CDM model and shows potential to alleviate the $S_8$ tension observed in weak lensing surveys.
However, the discrepancy with $f\sigma_8$ measurements indicates that the model may struggle to simultaneously account for the expansion history and the growth of structure.

It is important to note, though, that our analysis did not include weak lensing and $f\sigma_8$ data in the likelihood. A more comprehensive inference--carefully accounting for the fiducial cosmologies underlying these measurements--may yield improved consistency.
We leave such an analysis for future work.


\begin{acknowledgements}
We thank Jesus Torrado and David Camarena for useful comments and discussions.
VP thanks CAPES (Brazil) for financial support.
VM thanks CNPq (Brazil) and FAPES (Brazil) for partial financial support.
RvM is suported by Fundação de Amparo à Pesquisa do Estado da Bahia (FAPESB) grant TO APP0039/2023.
The authors would like to acknowledge the use of the computational resources provided by the Sci-Com Lab of the Department of Physics at UFES, which was funded by FAPES, CAPES and CNPq.
\end{acknowledgements}



\bibliographystyle{apsrev4-2}
\bibliography{biblio.bib}





\end{document}